\documentclass[12pt]{article}
\newcommand{\be}{\begin{equation}}
\newcommand{\ee}{\end{equation}}
\newcommand{\bea}{\begin{eqnarray}}
\newcommand{\eea}{\end{eqnarray}}

\title{A thermal instability for positive brane cosmological constant
in the Randall-Sundrum cosmologies}

\author{Stephon Alexander$^{2,3}$, Yi Ling$^{1,2\dag}$, Lee
Smolin$^{*1,2}$\thanks {email: s.alexander@ic.ac.uk, $^{\dag}$
ling@phys.psu.edu, $^{*}$
smolin@phys.psu.edu}\\\centerline{\footnotesize ${}^1$ Center for
Gravitational Physics and Geometry,  Department of Physics,}\\
\centerline{\footnotesize The Pennsylvania State University,
University Park, PA, USA 16802.}\\ \centerline{\footnotesize
${}^2$ The Blackett Laboratory, Imperial College of Science,
Technology and Medicine, London SW7 2BZ, UK}\\
\centerline{\footnotesize ${}^3$ Institute for Strings, Cosmology
and Astroparticle Physics, Columbia University, New York, NY
10027, USA}}
\begin{document}
\maketitle

\begin{abstract}
\baselineskip=20pt We describe a novel dynamical mechanism to
radiate away a positive four dimensional cosmological constant, in
the Randall-Sundrum cosmological scenario.  We show that there are
modes of the bulk gravitational field for which the brane is
effectively a mirror.  This will generally give rise to an
emission of thermal radiation from the brane into the bulk. The
temperature turns out to be nonvanishing only if the effective
four dimensional cosmological constant  is positive. In any theory
where the four dimensional vacuum energy is a function of physical
degrees of freedom, there is then a mechanism that radiates away
any positive four dimensional cosmological constant.
\end{abstract}

\section{Introduction}

\baselineskip=20pt In this letter we describe a new dynamical
mechanism, which, in the context of the Randall-Sundrum
cosmological scenario\cite{Ar,Ant,RS1,RS}, renders unstable
solutions described by the embedding of a four dimensional de
Sitter universe in the five dimensional $AdS$ spacetime. This
mechanism is closely related to the Unruh\cite{Unruh} effect and,
more directly, to the effect whereby accelerated mirrors
radiate\cite{mirror,BD}. When the four dimensional vacuum energy
(or cosmological constant) is a function of dynamical degrees of
freedom, such as when the vacuum energy depends on the value of a
scalar field, our result implies that positive vacuum energy
density may be radiated into the bulk, resulting in a decrease of
the effective four dimensional cosmological
constant\footnote{Earlier proposals for dynamical decay of the
cosmological constant in de Sitter spacetime were described in
\cite{early}.}.

More particularly, we describe here three results of an
investigation into the possible role of Unruh and moving mirror
acceleration in the context of the Randall-Sundrum cosmologies.

\begin{itemize}

    \item{}There are modes of the linearized bulk gravitational field
    which, rather than being
    bound to the brane, see the brane as a perfectly reflecting mirror.

    \item{}Given plausible physical assumptions, this implies that
    the brane does radiate into the $AdS$ bulk spacetime.  However,
    when the effective four dimensional cosmological
    constant $\lambda_{4e}$ vanishes, or is negative, the temperature
    is zero.  Thus, the original RS scenario in which $\lambda_{4e}=0$
is
    in fact stable and does not Unruh radiate into the bulk.

     \item{}When $\lambda_{4e}$ is strictly positive, the temperature
is
    non-vanishing.  The temperature is proportional to
    the square root of effective 4d cosmological constant,
    $\sqrt{\lambda_{4e}}$.

\end{itemize}

We note that so long as the approximations we employ are valid,
these results mean that any positive value of the cosmological
constant can be radiated away, with a temperature governed by
$\sqrt{\lambda_{4e}}$ \footnote{ Chamblin et al \cite{Chamblin},
previously discussed moving mirror radiation from branes and its
interaction with Hawking radiation from black holes in the bulk. A
related idea, studied previously by Deffayet et al\cite{Def} and
Levin\cite{Janna}, is that de Sitter radiation in the brane might
leak into the bulk, resulting in decay of the cosmological
constant.}.

In the next section we display the existence of linearized modes
of the bulk gravitational field which see the brane as a mirror.
In section 3 we apply to this known results on Unruh radiation in
$AdS$ spacetime and moving mirrors to deduce the conclusions just
mentioned. In the concluding section we discuss the range of
parameters in which our approximations may be considered valid.

\section{The Brane as a Mirror}

Randall and Sundrum and others have studied modes of the
linearized bulk Einstein equations which are, in the sense
described in \cite{RS1}, bound to the brane.  In this section we
demonstrate the existence of other linearized bulk modes, for
which the brane functions as a mirror.  We begin with the five
dimensional action for the RS scenario: \be S={1 \over G_5} \int
d^{5}x\sqrt{-g_{5}}({\cal R} \rm^{(5)}  - \Lambda  ) - \int
d^{4}x\sqrt{-g_{4}}(\tau + {\cal L}\rm_{matter}), \label{action}
\ee where ${\cal R} \rm^{(5)}$ is the five dimensional Ricci
scalar, $g_{5}$ and $g_{4}$ are the five and four dimensional
metrics respectively, $G_5= l_{Pl}^2 R$ is the five dimensional
Newton's constant and $\tau$ is the brane tension. The bulk space
is a piece of anti-de Sitter space described by \be ds^{2}=  dy^2
+ e^{-|y|/R} \eta_{\mu \nu} dx^\mu dx^\nu, \ee and the brane
resides at $y=0$.  We are interested in the boundary conditions
for gravitational modes at the brane. Hence, we begin by
linearizing the Einstein equations.

We begin by noting that it is possible to consider two classes of
bulk perturbations. The first are those identified by
Randall-Sundrum as those bound to the brane\cite{RS1}. \be
 ds^{2}=  dy^2 + e^{-|y|/R}[  \eta_{\mu \nu} + \gamma_{\mu \nu} ]dx^\mu
dx^\nu, \ee where $\gamma_{\mu \nu}$ is assumed to be  $C^\infty$
at the brane. This means that the derivatives of the linearized
modes actually have singular behavior near the brane, coming from
terms in $\partial_y |y| $ and $\partial_y^2 |y| $ at $y=0$. The
behavior of these modes is by now well studied to linearized
order. Let us call these type I modes.

These are not the only linearized modes of the metric in the
spacetime. Consider perturbations of the form, \be ds^2 =  dy^2 +
[e^{-|y|/R} \eta_{\mu \nu} + h_{\mu \nu} ]dx^\mu dx^\nu,
\label{II} \ee where now $h_{\mu\nu}(y,x)$ is considered to be
non-singular at the brane.  Let us call these type II modes. To
study them  it is more convenient to switch to coordinates
$z=Re^{y/R}$, for $y \geq 0$ (which implies that $z \geq R$), so
that the $AdS$ metric is of the form \be ds^{2}=
\frac{R^{2}}{z^{2}}(dz^{2} + dx_{4}^{2}), \ee and the linearized
metric is of the form, \be
 ds^{2}= \frac{R^{2}}{z^2} \left ( dz^{2} + dx_{4}^{2}  \right )
 + h_{\mu\nu}(z,x)  dx^\mu dx^\nu.
\ee Again $h_{\mu\nu}$ is assumed to be non-singular\footnote{Note
that while the background RS solution is invariant under a parity
transformation in which $y \rightarrow -y$, the same is not
required to be true of the linearization of solutions close to the
RS solution.  The point is that while a particular solution to
Einstein's equations may have an isometry, no isometry can be
imposed on the whole space of solutions to the Einstein's
equations. (In fact, only a set of measure zero of metrics in the
space of solutions have any isometries).  Thus, if we consider
linearizing the Einstein's equations around a given solution with
an isometry, the isometry cannot be imposed also on the linearized
modes.  The reason is that to be physically meaningful the
linearized modes must span the space of linearizations of exact
solutions close to the original one. Were we to impose a symmetry
of the background on the linearized modes we might miss some modes
which are linearizations of full solutions.  Thus we must consider
all type II modes which satisfy the boundary conditions, without
regard to symmetry.} at the position of the brane, which is now
$z=R$.

We choose a flat brane located at $z=z_{0}=R$ with $R^{2}=\Lambda$
and then linearize the field equations arising from the action
(\ref{action}).  It is sufficient for our purposes to go directly
to traceless, transverse gauge, where we find, \be
\frac{1}{\sqrt{-g_5}}\partial_{\rho}g_5^{\rho\sigma}
\partial_{\sigma}\sqrt{-g_5} h_{\mu\nu}= \tau \delta(z,z_{0})h_{\mu\nu}
+ T^{matter}_{\mu \nu}\delta(z,z_{0}). \ee Since the brane tension
$\tau$ is non-vanishing, all solutions to this equation in the
absence of matter on the brane must satisfy \be h_{\mu\nu}(z_{0})=
0. \ee This is simply the Dirichlet boundary condition. Thus, we
find that type II modes which satisfy the linearized Einstein's
equations exist, but they are not free on the brane, instead, in
the absence of matter, the brane acts as a mirror for these
modes\footnote{Note that for Type I modes the dangerous $\tau
\delta(z,z_0)$ term is canceled by a delta function coming from
the term $\partial^2_y |y|/R$ which comes from the five
dimensional wave operator applied to the perturbation
$e^{-|y|/R}h_{\mu \nu}$. }.

This result is not special to the flat embedding, it applies to
all fluctuations around brane embeddings which extremize the
action (\ref{action}) of the form, \be ds^2 =  dy^2 + [ g^0_{\mu
\nu} + h_{\mu \nu} ]dx^\mu dx^\nu, \label{IIgeneral} \ee where
$g^0_{\mu \nu}$ is the metric for an embedding of a brane.

Type II modes were considered also by Randall and
Sundrum\cite{RS1}, who argued that they interact very weakly with
matter on the brane. While this is true, we will show below that
the fact that they see the brane as a mirror has consequences due
to the quantum effect that accelerated mirrors in some
circumstances radiate.

Before going on to understand what the consequences of the type II
modes are for the RS cosmological scenarios we should comment on
the existence of two kinds of modes of the linearized metric, Type
I, bound to the brane and Type II, which are reflected by the
brane. How can we tell if either, or both are true modes of the
gravitational field?  The point is that not any solution to the
linearized Einstein's equation represents an actual linearization
of a full solution to the full Einstein's equations. What is
required is more\cite{wald}: $h_{\mu \nu}$ is a genuine
linearization of a full solution to Einstein's equations if there
is a one parameter family of exact solutions $g_{\mu
\nu}(\lambda), \phi(\lambda )$, where $\phi$ represents all the
matter degrees of freedom, such that $dg_{\mu \nu}(\lambda)
/d\lambda|_{\lambda =0} =h_{\mu \nu}$. One simple check is to
carry out an expansion of the Einstein equations to higher order,
to show that a solution can be generated order by order. In the
case of Type I modes these equations may be singular as a result
of the $\partial_y |y| \approx \theta (y) $ and $\partial_y^2 |y|
\approx \delta (y)$  behavior of the derivatives of the modes at
the location of the brane at $y=0$. This is a question that
deserves investigation.

\section{The Brane Mirror and Thermal Radiation}

The problem of moving mirror boundary conditions in quantum field
theory has been studied in some detail since the
1970's\cite{mirror,BD}. The conclusion in Minkowski spacetime is
that a moving mirror with a constant acceleration will radiate a
thermal bath of particles with a temperature given by the Unruh
radiation associated to its value of acceleration, $a$ \be
T_{mm}=T_{unruh}= {\hbar a \over 2\pi c}. \ee It is important to
note that this is a nontrivial observation, as a moving mirror is
a different physical system than the detectors first studied by
Unruh and others.  However it is easy to make a thermodynamic
argument for it. If I am in a bath of thermal radiation at
temperature $T$ and I uncover a mirror, or take one out of my
pocket, that mirror will interact with the radiation and come to
equilibrium at the same temperature. Any other behavior would
contradict the second law of thermodynamics. But since an
accelerating observer in the vacuum is known to observe a thermal
bath of radiation, the same must apply to this case, otherwise an
accelerated observer could violate the laws of thermodynamics.

The problem of Unruh radiation has also been studied in $AdS$
spacetimes, by Deser and Levin\cite{DL1,DL2}, and
Jacobson\cite{TedAdS}, following a seminal work on $QFT$ in $AdS$
spacetimes by Avis, Isham and Storey\cite{Avis}. They find that
there are several natural choices of vacua, corresponding to the
fact that $AdS$ spacetime is not globally hyperbolic.  The
specification of the vacuum state then depends on boundary
conditions imposed at timelike infinity.  Isham et al, and Deser
and Levin study three natural choices of vacua, corresponding to
reflected or transmitted boundary conditions at the time like
infinity. They find that a particle detector following a timelike
worldline in a $d$ spacetime dimensional $AdS$ spacetime will in
some cases detect a flux of particles with a thermal or modified
thermal spectrum.  They further find that there are simple
criteria to determine whether the temperature is non-vanishing,
and compute its value.  These  have to do with the embedding of
the accelerating trajectory in the flat $d+1$ dimensional
spacetime within which the $AdS$ spacetime is embedded.  First,
the temperature is only non-vanishing when the trajectory in the
$d+1$ dimensional spacetime is hyperbolic, so it has horizons in
the embedding spacetime, analogous to the Rindler horizon.  Then,
when the temperature is non-vanishing it is in fact proportional
to the $d+1$ dimensional acceleration of the worldline in the
$d+1$ dimensional flat embedding spacetime. \be T_{unruh} ={ \hbar
a_{d+1} \over 2 \pi c}. \label{acc} \ee

The moving mirror problem has not, to our knowledge, been studied
in an $AdS$ spacetime. But by making a similar thermodynamic
argument we can arrive at the conclusion that a moving mirror in
$AdS$ spacetime will radiate quanta in a thermal or modified
thermal spectrum at a temperature given by (\ref{acc}).  Suppose a
family of observers in ordinary $AdS$ spacetime follows the
trajectories that would be followed by freely falling worldlines
of an RS brane.  By the results of \cite{DL1,DL2,TedAdS} they
would see themselves immersed in a bath of radiation with
temperature (\ref{acc}). Now suppose they convert some sand they
are carrying into mirrors that totally reflect the quanta in
question. Then those mirrors must also  come to equilibrium at the
same temperature (\ref{acc}), otherwise the second law is
violated. Thus, an external observer, moving inertially in the
$AdS$ spacetime will see the mirrors radiate a flux of quanta at
the same temperature.

Of course the energy must come from the observers, who require
energy to both continue their motion and construct the mirrors.
The energy drained by the radiation created represents a kind of
purely quantum mechanical radiation reaction.  In the case that
the brane itself acts as a mirror, the energy must come from
energy density stored in the vacuum of the quantum field theory on
the brane.

In the previous section we showed that the RS brane is precisely a
mirror for a certain set of modes in the bulk five dimensional
$AdS$ spacetime, which we called Type II modes. We then have a
prediction that the RS brane will radiate into the bulk a gas of
thermal Type II gravitons \be T_{RS}= { \hbar a_{6} \over 2 \pi
c}, \label{a6} \ee where $a_6$ is the acceleration of a freely
falling particle in the RS brane measured in the six dimensional
embedding spacetime.

We need now only look at the details to find out what this implies
for RS cosmologies.

The RS brane is embedded in $AdS_5$ which is an embedding into a
flat six dimensional space \be
ds^2_6=-d\tilde{y}^2_0+d\tilde{y}^2_1+d\tilde{y}^2_2+
d\tilde{y}^2_3+d\tilde{y}^2_4-d\tilde{y}^2_5, \ee with constraint
\be -\tilde{y}_{0}^2 + \tilde{y}_{1}^{2} +\tilde{y}_{2}^{2} +
\tilde{y}_{3}^{2} + \tilde{y}_{4}^{2} - \tilde{y}_{5}^{2}=-R^2.
\ee The coordinates can be parameterized by defining \bea
\tilde{y}_0&=&{1\over 2z}(z^2+x_i^2-t^2+R^2),\nonumber\\
\tilde{y}_i&=&{Rx_i \over z},\nonumber\\
 \tilde{y}_4&=&-{1\over 2z}(z^2+x_i^2-t^2-R^2),\nonumber\\
\tilde{y}_5&=&{Rt\over z},\label{ct1} \eea such that the metric
can be written as, \be ds^2_5={R^2\over z^2}(dz^2+dx_i^2-dt^2),
\ee where index $i$ runs from 1 to 3. If we define, \be
z=Re^{y/R}, \label{ct2} \ee then it takes the form, \be
ds^2_5=dy^2+e^{-2y/R}(dx_i^2-dt^2). \label{min} \ee

To consider the $dS$ embedding, let us introduce a positive
constant $\Lambda_4$, and take the following coordinate
transformation\cite{Kal,De,KR} \bea y &=& -\sqrt{\Lambda_4} Rt'-
Rlog(sinh  \frac{ \gamma - y'}{R} ), \nonumber \\ t&=&-R coth
\frac{\gamma - y'}{R}e^{-\sqrt{\Lambda_4}t'}, \nonumber \\
x_i&=&\Lambda_4R^2 x_i'. \label{ct3} \eea

The metric becomes \be ds^2_5=d^2y'+(\Lambda_4 R^2sin^2h {{\gamma
- y'}\over R})(e^{2\sqrt{\Lambda_4}t'}d^2x_i'-d^2t'),
\label{dsbrane} \ee

where $\gamma$ is related to the brane tension by, \be
\tau=\frac{3}{G_5 R}coth \frac{\gamma}{R}. \ee It's evident that,
at $y'=const.$, the metric of the brane describes the standard $de
Sitter$ spacetime\footnote{ We emphasize that equation
(\ref{dsbrane}) is indeed a solution to the action (\ref{action}).
For details, see \cite{Kal,KR}; for de Sitter embeddings with two
rather than one branes, see \cite{De}.}. We also see that a de
Sitter embedding is only possible for $\tau > 3/G_5 R$. However as
can be seen from (\ref{4dcos}), below, this does not restrict the
effective four dimensional cosmological constant.

Now suppose a graviton detector is set at $x_i=0$ on the brane.
Despite being static with respect to brane world, it is
accelerated in both $AdS_5$ and the six-dimensional flat space
time. Furthermore, the trajectory of the detector can be
identified with the trajectory of the brane in $AdS_5$ or $M_6$.
We are particularly interested in the six-dimensional acceleration
$a_6$. When $y'=const.$ and $x_i=x^{'}_{i}=\tilde{y}_{i}=0$, \be
-\tilde{y}_0^2+\tilde{y}_4^2=R^2 cosh^2 {{\gamma - y'}\over
R}-R^2\equiv a_6^{-2}, \ee where $a_6$ is the constant
acceleration in six dimensional flat space time. We see that the
trajectory is indeed hyperbolic in the six dimensional spacetime,
which means that the temperature will be non-zero. We can also
compute the magnitude of the six dimensional acceleration directly
by making use of equations (\ref{ct1}), (\ref{ct2}) and
(\ref{ct3}) to compute $d^2\tilde{y}^\alpha /ds^2$, where $s$ is
the proper time of the detector. As $x'_i=dx'_i=0$ and
$y'=const.$, $s=\sqrt{\Lambda_4} R sinh\frac{\gamma -y'}{R}t'$. It
is straightforward to obtain the magnitude of acceleration which
is, \be |a_6|^2=-|a_{y_0}|^2+|a_{y_4}|^2=\frac{1}{R^2
sinh^2\frac{\gamma - y'}{R}}. \ee Thus, by combining this with the
results of \cite{DL1,DL2} we conclude that an observer at
$y'=const.$ and $x_i=x'_{i}=y_{i}=0$ on a de Sitter brane will
detect thermal radiation with the temperature \be T={\hbar
a_{6}\over 2\pi c} = \frac{\hbar}{2\pi c R sinh{\frac{\gamma -
y'}{R}}}. \label{main} \ee To find the effective four dimensional
cosmological constant $\lambda_{4e}$ we must renormalize the
coordinates so that the metric is in the form of \be
ds^2_{brane}=e^{2\sqrt{\lambda_{4e}(y')} s}d^2w_i-d^2s.
\label{dsbrane2} \ee This requires new coordinates, \be ds=
(\sqrt{\Lambda_4} R sinh {{\gamma- y'}\over R}) dt', \ee \be dw_i=
(\sqrt{\Lambda_4} R sinh {{\gamma- y'}\over R}) dx_i'. \ee The
effective cosmological constant for a brane positioned at $y'$ is
then \be \lambda_{4e}={1 \over {R^2 sin^2h{{\gamma- y'}\over R}}}.
\label{4dcos} \ee We see then that at fixed $y'$ the temperature
is related to the effective cosmological constant by the usual de
Sitter relation, \be
    T=\frac{\hbar\sqrt{\lambda_{4e}}}{2\pi c}.
    \label{right}
\ee

By the above thermodynamic argument we then expect that the brane
itself will radiate into the modes for which it serves as a
mirror, with the same temperature.  These are our main results.

We now make several comments on these results.

\begin{itemize}

    \item{}We note that (\ref{right}) is the same as the de Sitter
    temperature of de Sitter radiation internal to the brane, which
    is a consequence of the existence of horizons in the de Sitter
    spacetime.  However the moving mirror radiation radiates
    into the five dimensional $AdS$ spacetime. As the brane is the
    source of that radiation it must result in energy loss from the
    brane.

    \item{}If we consider the Minkowski and $AdS$ embeddings, which
    have vanishing or negative cosmological constant, respectively, we
can easily show that there is no thermal radiation in these cases.
To see this, we consider the flat brane which is obtained by
fixing $y$ to be constant in (\ref{min}), and a quick calculation
shows that the acceleration in six dimensional spacetime is zero
\footnote{Even so, note the five dimensional acceleration over the
$AdS$ background is not zero. The brane is accelerated with
respect to the bulk at the rate of $1/z_0$, where $z_0(\le R)$ is
the transverse position of the brane.}, implying a vanishing
temperature. As far as the $AdS$ brane is concerned, we have the
following choice\cite{De}, \bea y &=& -\sqrt{-\Lambda_4} Rx'_3-
Rlog(cosh  \frac{ \gamma - y'}{R} ), \nonumber \\ x_3&=&R tanh
\frac{\gamma - y'}{R}e^{-\sqrt{-\Lambda_4}x'_3}, \nonumber
\\
x_{1,2}&=&-\Lambda_4R^2 x'_{1,2}, \ \ \ \ \ \ \ \ \ t =
-\Lambda_4R^2t', \label{ct4} \eea

such that the metric has the form, \be ds^2_5=d^2y'+(-\Lambda_4
R^2cosh^2 {{\gamma - y'}\over R}) [e^{-2\sqrt{-\Lambda_4}x'_3}
(dx_{1}^{\prime 2}+dx_{2}^{\prime 2}-dt^{\prime 2})+
dx_{3}^{\prime 2}]. \ee

From (\ref{ct1}), (\ref{ct2}) and (\ref{ct4}), we find the
six-dimensional acceleration of a  detector at a fixed spatial
position on the brane is also zero. Thus, the temperature of the
brane is given by \be T={ \hbar a_{6}\over 2\pi c} = { \hbar
\Theta(\lambda_{4e})\sqrt{\lambda_{4e}} \over {2\pi c }}. \ee

\item{}It of course follows from (\ref{main})
that, if the effective cosmological constant of our universe is
positive, all the fields we observe are very weakly coupled to a
five dimensional thermal bath of Type II gravitons with a
temperature  given by (\ref{right}).   Of course, the coupling is
extremely weak, as has been discussed by \cite{RS1} and other
authors. But it is not impossible that there may be observable
consequences of this prediction.

\end{itemize}

\section{Conclusion: A Possible Dynamical Mechanism for Radiating
away Positive Cosmological Constant}

The results just describe imply that the embedding of a four
dimensional de Sitter universe as a brane in the five dimensional
Randall-Sundrum cosmology is unstable.  The exact channel that the
instability goes through depends on the coupling of the quantum
fluctuations of the vacuum to the Type II modes and hence on the
model used for the vacuum energy.  Of course, as noted in
\cite{RS1} the coupling of matter degrees of freedom living on the
brane to the Type II modes may be suppressed by various factors.
Let us first consider the case where the suppression is
insignificant, so that the radiation may be considered to be
blackbody. In this case the energy radiated into the bulk per unit
time and unit brane volume is given by the usual formula, \be
{\cal R}= c \hbar^{-4}  T^5  \approx  \lambda_{4e}^{5/2}. \ee It
is not difficult to see that this effect may be significant for
the evolution of the early universe. Let us consider a standard
inflationary model in which $\tau = {\cal V}(\phi )$ where $\phi$
is the inflation field and ${\cal V}(\phi )$ its potential.  Let
$\lambda_{4e} =1/L^2$ and define the evaporation time scale by \be
t_{evap} = \left ( {1 \over \tau } {d\tau \over ds} \right )^{-1}
. \ee Then an adiabatic approximation,  in which the solution can
be approximated by a series of de Sitter embeddings with slowly
decreasing $\lambda_{4e}$, will be valid so long as \be {t_{evap}
\over L} >> 1 . \ee But it is then not difficult to see that \be
{t_{evap} \over L} > 3 \left ( {L \over l_{Pl}} \right )^2 \left (
{L \over R} \right )^2 \ee so that if $R \approx Tev^{-1}$ to
solve the hierarchy problem then an adiabatic approximation is
good so long as $L > 10^{7} l_{Pl}$. At the same time, we see that
the evolution of an inflationary solution may be significantly
modified at early times.

The rate of decay may be slowed by suppression factors coming from
the fact that the coupling of Type II modes to the vacuum energy
is weak and gravitational. But the effect of these will be to
increase the evaporation time, which increases the regime in which
an adiabatic approximation is valid.

We note that the effect is present so long as $\lambda_{4e} >0$,
which means that $\tau > 3/G_5 R$.  Thus this effect may radiate
away all contributions to the effective cosmological constant,
including those that result from matter falling onto the brane
from the bulk, such as described in  (\cite{carsten, Carsten}). In
principle, the radiation continues until $\lambda_{4e} =0$, at
which point $\tau $ decreases to the value necessary to stabilize
the embedding of the flat $RS$ brane. However the evaporation time
becomes much longer than $L$  as $\lambda_{4e}$ decreases. More
work is then needed to understand whether this effect may be
significant for observational astronomy.

In closing we remark on several checks of our reasoning that
should be carried out.  First, the thermodynamic argument we rely
on here should be checked by doing a first principles calculation
of radiation from a moving mirror in a five dimensional $AdS$
spacetime. These calculations are in progress\cite{yi}. Given
these results one may investigate, in the context of particular
models of the inflation, whether the effective cosmological
constant is in fact sent to zero and how the transition from an
inflating universe to a FRW universe takes place.  Finally, the
question of whether both Type I and Type II modes are genuine
linearizations of full solutions to the Einstein equations should
be investigated by carrying out the linearizations to higher
order.

\section*{Acknowledgments}

We are grateful to Daniel Freedman, Kelle Stelle, Emmanual Katz,
Janna Levin and Lisa Randall for discussion and correspondence and
to Clifford Burgess, Robert Brandenberger, Andrew Chamblin, Gia
Dvali, Ted Jacobson, Nemanja Kaloper  and Robert Myers for
comments on earlier versions of this letter. Yi Ling and Lee
Smolin would like to thank Chris Isham and the theoretical physics
group at Imperial College for hospitality. This work was supported
by the NSF through grant PHY95-14240, a PPARC grant and gifts from
the Jesse Phillips Foundation.

\end{document}